\documentclass[twocolumn,5p]{elsarticle}
\usepackage{amsmath}
\usepackage[utf8]{inputenc}
\usepackage{graphicx}
\graphicspath{ {./images/} }
\usepackage{float}
\usepackage{subcaption}
\PassOptionsToPackage{hyphens}{url}
\usepackage{hyperref}
\biboptions{numbers,sort&compress}

\begin{document}
\begin{frontmatter}

\title{The electronic interface for quantum processors}

\author[tud,kav]{J.P.G.~van Dijk}
\author[tud,kav,intel,epfl]{E.~Charbon}
\author[tud]{F.~Sebastiano\corref{cor1}}
\ead{f.sebastiano@tudelft.nl}
\cortext[cor1]{Corresponding author}
\address[tud]{QuTech, Delft University of Technology, P.O. Box 5046, 2600 GA Delft, The Netherlands}
\address[kav]{Kavli Institute of Nanoscience, P.O. Box 5046, 2600 GA Delft, The Netherlands}
\address[intel]{Intel Corporation, 2501 NW 229th Ave, Hillsboro, OR 97124, USA}
\address[epfl]{\'{E}cole Polytechnique F\'{e}d\'{e}rale de Lausanne, Case postale 526, CH-2002 Neuch\^{a}tel, Switzerland}

\date{\today}

\begin{abstract}
Quantum computers can potentially provide an unprecedented speed-up with respect to traditional computers. However, a significant increase in the number of quantum bits (qubits) and their performance is required to demonstrate such \emph{quantum supremacy}. While scaling up the underlying quantum processor is extremely challenging, building the electronics required to interface such large-scale processor is just as relevant and arduous. This paper discusses the challenges in designing a scalable electronic interface for quantum processors. To that end, we discuss the requirements dictated by different qubit technologies and present existing implementations of the electronic interface. The limitations in scaling up such state-of-the-art implementations are analyzed, and possible solutions to overcome those hurdles are reviewed. The benefits offered by operating the electronic interface at cryogenic temperatures in close proximity to the low-temperature qubits are discussed. Although several significant challenges must still be faced by researchers in the field of cryogenic control for quantum processors, a cryogenic electronic interface appears the viable solution to enable large-scale quantum computers able to address world-changing computational problems.
\end{abstract}


\end{frontmatter}

\section{Introduction}

Quantum computers hold the promise to ignite the next technological revolution as the classical computer did for last century’s digital revolution, both by efficiently solving problems that are intractable even for today’s supercomputers, such as simulation of quantum systems, and by accelerating classical applications. For instance, by enabling the efficient simulation of quantum systems, quantum computing could help in synthesizing a room-temperature superconductor, which would significantly reduce energy loss in power lines, generators, and supercomputers \cite{svore2016quantum}. It could also contribute to improving the efficiency of industrial nitrogen fixation into fertilizers, which is currently achieved through the 117-year-old Haber process and uses up to 5\% of the natural gas produced each year worldwide \cite{gibney2014physics,reiher2017elucidating}. Furthermore, quantum computers have the potential to solve classical problems with much higher efficiency, such as searching in huge datasets, solving large linear systems, and even addressing privacy concerns by running algorithms without the operator being able to detect what is being computed \cite{svore2016quantum}. The ability to solve currently intractable problems and significantly accelerate certain computations represents a game changer with the potential to revolutionize entire industries.

\begin{figure}[ht]
\centering
\includegraphics[width=1\linewidth]{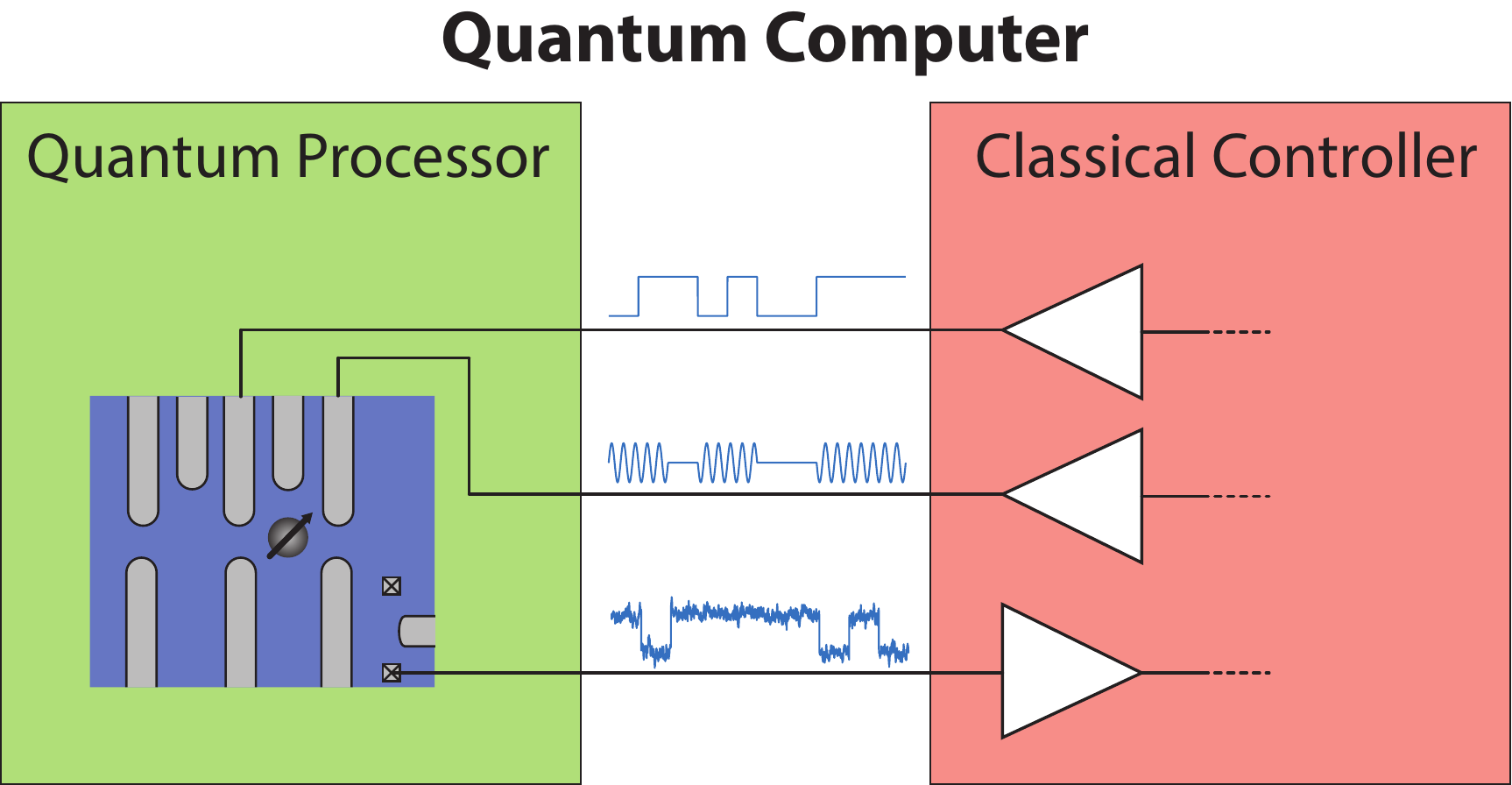}
\caption{\label{fig:quantum_computer}A quantum computer comprises a quantum processor and a classical electronic controller that generates and processes the various signals.}
\end{figure}

\begin{figure*}[ht]
    \centering
    \begin{subfigure}[b]{0.45\textwidth}
        \centering
        \includegraphics[height=4.25cm]{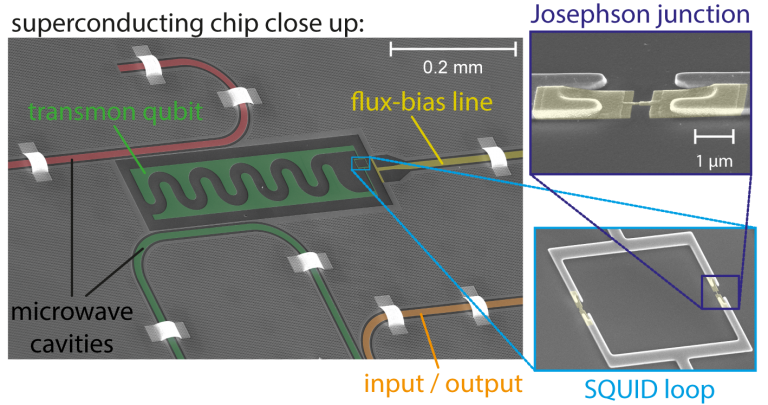}
        \caption{\label{fig:tech_super}A typical transmon qubit, composed of a capacitor in parallel with a SQUID loop comprising two Josephson junctions; a flux-bias line to tune the qubit frequency and microwave cavities used for control and read-out are also shown \cite{langford2017experimentally} (image taken from Ref. \cite{superconductingqubitimage}).\\}
        \label{fig:gull}
    \end{subfigure}
    \begin{subfigure}[b]{0.5\textwidth}
        \centering
        \includegraphics[height=4.25cm]{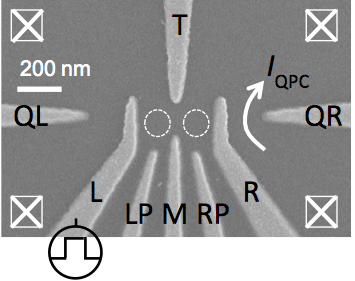}
        \caption{\label{fig:tech_semi}A typical semiconductor qubit: two quantum dots (dotted circles) are formed by a set of metallic electrodes depleting the 2D electron gas (2DEG); electrons can be loaded from the reservoir connected by ohmic contacts (crossed squares); a quantum point contact (QPC) for the read-out and its current ($I_{QPC}$) is present on the right side (image taken from Ref. \cite{kim2014}).}
        \label{fig:tiger}
    \end{subfigure}
    \caption{\label{fig:tech_pics}Pictures of the actual quantum processor chips in different solid-state quantum technologies.}
\end{figure*}

To achieve such tremendous progress, a quantum computer relies on processing the information stored in quantum bits (qubits) that must be typically cooled at cryogenic temperatures in dilution refrigerators for proper operation. Nevertheless, a quantum computer is more than only a quantum processor hosting several qubits, since all operations must be controlled by a classical (i.e.~non-quantum) electronic interface, as shown in Fig.~\ref{fig:quantum_computer}. Today, such an electronic interface can be implemented by off-the-shelf electronic components and instrumentation, as state-of-the-art quantum processors comprise only a few qubits (less than 20 \cite{IBM17qubits,kelly2015state,monz201114}), thus showing very limited complexity in the number of subsystems. However, quantum processors comprising thousands or even millions of qubits will be required to address any of the practical computing problems mentioned before  \cite{fowler2012surface}. Although building such a large-scale quantum processor is a daunting task by itself, realizing a large-scale electronic infrastructure to handle a very large number of qubits is also a formidable challenge. Due to the limited size of existing quantum processors, the electronic interface is not yet the main limitation in the performance of quantum computers, but it could become a major bottleneck in the near future, as both the performance and the complexity of quantum processors advance.

To address this challenge, this paper presents the obstacles to be faced when scaling up a quantum computer, with specific focus on the electronic interface connecting to the quantum processor. Section \ref{sec:requirements} discusses the requirements of such electronics when employed in combination with processors adopting different qubit technologies. Since we focus on the need for large-scale quantum computation, the scope will be limited to the qubit technologies that today promise large-scale integration, i.e.~solid-state qubits requiring only purely electrical control, such as spin qubits and superconducting qubits. An overview on how the requirements dictated by the quantum processor are handled by state-of-the-art electronic controllers is given in Section \ref{sec:controllers}. Then, the challenges in scaling up the existing controllers and possible approaches to tackle them will be treated in Section \ref{sec:mux} and \ref{sec:scalable_controller}, respectively.
Conclusions are drawn in Section \ref{sec:conclusions}
 \section{\label{sec:requirements}Quantum Processor Requirements}
DiVincenzo defined five criteria that a quantum processor must satisfy to be employed in a quantum computer \cite{divincenzo2000physical}. In short, the quantum processor must be scalable and contain well characterized qubits (I) that can be initialized (II) and can preserve their quantum state for a time\footnote{The coherence time is the characteristic time constant of the process of a qubit losing its relevant quantum behavior  due to unwanted interactions with the environment.}  ($T_2^*$) sufficiently longer than the duration of a quantum operation (III). Finally, it must be possible  to operate on the qubits with a universal set of quantum gates, i.e.~both single-qubit and two-qubit operations, (IV), and to read out the state of a qubit (V).

This section will give a brief overview of the implementation of the read-out, the single-qubit and the two-qubit gates in different quantum technologies, and the implications for the electronic controller.
As the state of a qubit can be represented geometrically as a point on a sphere and single-qubit operations correspond to rotations along this sphere, these operations will also be referred to (here and in the literature) as rotations in the following\footnote{To satisfy DiVincenzo's criteria, it is in general sufficient to be able to rotate the qubit state around two axis, typically the X and Z axis.}.
Specific attention will be devoted to assess the performance of each technology in terms of coherence time, and to analyze their potentials and hurdles for scalability. The technologies under consideration are superconducting qubits (Fig. \ref{fig:tech_super}), specifically transmons, and spin qubits in semiconductors (Fig. \ref{fig:tech_semi}), specifically single-electron spin-qubits (in quantum dots or bound to donors) and multiple-electron qubits, such as singlet-triplet qubits, exchange-only qubits and hybrid qubits. Typical values and control methods as used in current experiments are reported, but these may shift in the future as the various quantum technologies develop.

\subsection{Superconducting Qubits}

Various superconducting-qubit implementations have been proposed based on encoding the quantum information in the charge, phase or flux state of a superconducting resonator. However, the research community has generally converged to the adoption of the \emph{transmon}, a superconducting charge qubit with reduced sensitivity to charge noise, leading to improved coherence times \cite{koch2007charge,dicarlo2009, barends2014, chow2014, jeffrey2014, riste2015,reagor2018demonstration}, e.g.~\mbox{$T_2^*$ $\sim$ 18.7 $\mu$s} \cite{reagor2018demonstration}

In a transmon, the qubit states are defined as the two lowest energy levels of a non-linear resonator implemented using Josephson junctions in a loop shunted by a capacitor (Fig.~\ref{fig:tech_super}). The magnetic flux coupled to the loop of Josephson junctions can be controlled by means of a current, called the flux bias, running in close proximity. This allows for the qubit energy levels, and hence the qubit frequency, to be tuned externally. To allow for rapid changes of the qubit frequency, the flux-bias lines are typically implemented as transmission lines with short-circuit termination.

Transmons can be fabricated as integrated circuits by depositing and patterning a superconducting film and integrating Josephson junctions on a solid-state substrate, typically made out of silicon or sapphire. Control and read-out is performed  by coupling the qubit to microwave cavities on the same chip. Single-qubit rotations that flip the qubit state (X-rotation) can be obtained by applying  a microwave pulse  with  frequency matching the qubit resonance, typically in the 4-6-GHz range \cite{riste2015,barends2014}. Microwave pulses with Gaussian envelopes and in/quadrature-phase (I/Q) modulation adopting the Derivative Reduction by Adiabatic Gate (DRAG) approach \cite{motzoi2009simple}  are used to minimize leakage to higher energy states of the resonator. Typical parameters for the Gaussian envelope of the in-phase component are a 5-ns $\sigma$ and 20-ns total duration for a $\pi$-rotation \cite{riste2015}. As an alternative to a Z-rotation in software \cite{mckay2017efficient}, a single-qubit rotation around the Z-axis can be obtained by temporarily detuning the qubit frequency by adapting the flux bias \cite{dicarlo2009}.

A two-qubit gate can be obtained between two qubits coupled to the same microwave cavity. The cavity is off-resonance with both qubits during normal operation, but, by detuning the qubit frequencies via their flux biases, the cavity can couple to the qubits by virtual photon exchange \cite{dicarlo2009}. In this way, by applying proper pulses to the flux-bias lines of the two qubits, either a SWAP gate or a CPHASE gate can be implemented, depending on whether the flux bias is pulsed suddenly (typ. 12 ns \cite{riste2015}) or fast but adiabatically (typ. 40 ns \cite{dicarlo2009, barends2014, riste2015}), respectively. Alternatively, a microwave-controlled two-qubit CPHASE gate can be engineered \cite{chow2011simple, chow2013microwave}.

The qubit state is read out by probing the frequency response of a microwave cavity coupled to the qubit. A frequency shift of the cavity resonance frequency dependent on the qubit state can be observed in the measurement of the transmission coefficient of the cavity.  The microwave excitation used for the measurement has a typical frequency in the range of 7-8 GHz with a duration of $\sim$1 $\mu$s \cite{riste2015}. 

Superconducting qubits can be reliably fabricated nowadays, having led to quantum processors containing up to 72 qubits \cite{google72}. However, the individual qubits are currently very large (in the order of 0.1-1 mm$^2$ per qubit), which will hinder further scaling to thousands and millions of qubits per quantum processor. Furthermore, superconducting qubits have to operate at extremely low temperatures (typically $<$ 100 mK).

\subsection{Single-electron Semiconductor Qubits}
Semiconductor qubits are significantly smaller than transmons, with a typical pitch between neighbouring qubits in the order of 100 nm, therefore representing promising candidates for large-scale quantum processors. As charge qubits in semiconductors show poor coherence times due to their sensitivity to charge noise, most research is currently focused on spin qubits, in which the quantum state is encoded in the spin of one or more particles, such as nuclei and electrons, trapped in a specific location in a semiconductor solid-state substrate. A large magnetic field in the order of one Tesla is required to obtain a sufficiently large energy difference between the spin-up and spin-down states. The energy splitting between those two states determines the resonant frequency of the qubit. While such a magnetic field can be relatively easily generated and applied, it becomes challenging when a field gradient is desired over many qubits on the same chip to create addressibility \cite{kawakami2014, watson2018}. The decoherence of spin qubits is typically limited by the interaction with nuclear spins of all the other atoms in the substrate, which in turn can be minimized by using isotopically purified silicon \cite{veldhorst2014}. As a major advantage, semiconductor qubits  can be fabricated using industry-standard CMOS fabrication processes \cite{maurand2016cmos}, thus leveraging the very-large-scale of integration (VLSI) and the yield and the reproducibility of the semiconductor industry.   An additional advantage is that they can potentially be operated at higher temperatures, up to 4 K, although this has not yet been demonstrated \cite{vandersypen2016,petit2018spin}. However, the required small qubit pitch poses a challenge in routing all control lines individually.

The simplest spin qubit uses the spin of a single electron trapped in a quantum dot \cite{nowack2011, kawakami2014, veldhorst2014, veldhorst2015, watson2018} with the qubit states encoded in the spin-up and spin-down states, as initially proposed by Loss and DiVincenzo \cite{loss1998quantum}. A quantum dot hosting single electrons can be created by a set of gate electrodes locally forming an island in a 2-dimensional electron gas (2DEG), as shown in Fig.~\ref{fig:tiger}. Coherence times of  $T_2^*$ = 1 $\mu$s \cite{watson2018} and $T_2^*$ = 120 $\mu$s \cite{veldhorst2014} can be achieved in natural silicon and isotopically purified silicon, respectively.

Single-qubit rotations across the X-axis can be obtained by applying a varying magnetic field interacting with the magnetic dipole of the electron. Such a field can be simply generated by  a current running in a  transmission line nearby the qubit \cite{veldhorst2014}. Alternatively, by applying a microwave excitation to a gate that is capacitively coupled to the quantum dot, the electron can be forced to oscillate in a magnetic-field gradient generated by a local magnet \cite{kawakami2014}. The required microwave frequency must match the energy difference between the two spin states, and is typically in the range of 13-40 GHz with gate times of 0.2-2 $\mu$s for a $\pi$-rotation \cite{kawakami2014, veldhorst2014, watson2018}.

\begin{table*}[ht]
\centering
\caption{\label{tab:technologies}Summary of the various signals required to implement the various gate operations in different qubit technologies.}
\includegraphics[width=0.8\linewidth]{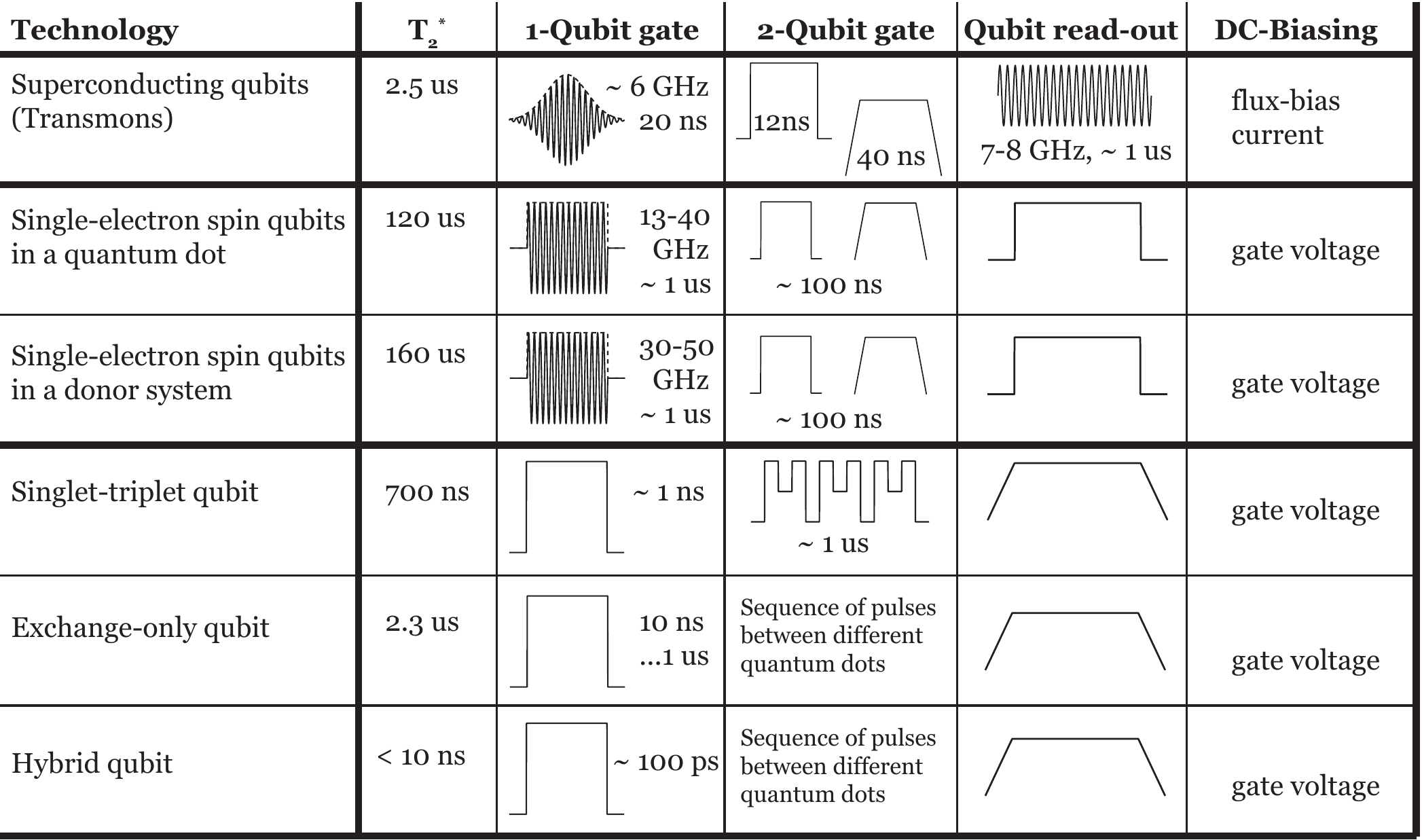} \end{table*}

Electrons in neighbouring quantum dots must be isolated by tunnel barriers during normal operation. For two-qubit operation, the voltages coupled to the quantum dots or to the tunnel barriers can be adapted to allow interaction between two neighbouring electrons, thereby implementing either a SWAP gate or a CPHASE gate  depending on whether the pulses applied to the gates are sudden or adiabatic. First demonstrations used exchange couplings around 3...10 MHz \cite{veldhorst2015, watson2018}, leading to pulse durations $\sim$100 ns.

Read-out is typically performed by converting information on the electron spin into the position of the electron and sensing its charge. For this spin-to-charge conversion, energy-selective tunneling to an electron reservoir \cite{elzerman2004single} or Pauli-exclusion tunneling to another dot \cite{ono2002current}  can be used. To detect the charge, either a charge sensor, such as a quantum point contact (QPC) or single-electron transistor (SET), or gate-based dispersive read-out \cite{colless2013}  can be employed. In both cases, the read-out involves measuring an electrical impedance: for charge sensors, a variation of a few percents in a resistive impedance ($\sim$25 k$\Omega$ for QPCs, $\sim$100 k$\Omega$ for SETs); for gate-based read-out, a variation of $\ll$ 1 fF on a capacitive impedance in the order of 1 pF.

Instead of trapping an electron in a quantum dot, an electron bound to a donor in the semiconductor substrate can be used, e.g.~the excess electron provided by a phosphorous dopant atom in silicon \cite{pla2012, pla2013, kalra2014, muhonen2014}. While accurately placing the donor atoms poses a fabrication challenge, such a system has the advantage that, unlike quantum dots, all atoms are identical, thus allowing for reproducible behavior. In addition, the atom nuclear spin can be used as quantum memory for long-term storage of a quantum state ($T_{2,nuc}$ = 30 s has been demonstrated \cite{muhonen2014}). A typical donor-based system shows coherence times of $T_2^*$ = 55 ns \cite{pla2012}, and $T_2^*$ = 160 $\mu$s \cite{muhonen2014} in purified silicon.
The operations on the donor-bound electron are implemented in the same way as for the electron trapped in a quantum dot. While the single-qubit operation times are similar, the qubit frequency in a donor system is typically slightly higher, i.e.~in the range of 30-50 GHz \cite{pla2012, pla2013, muhonen2014}. Two-qubit gates have not yet been published, since they would require the donors to be placed closer than 20 nm \cite{vandersypen2016}. This is currently hard to fabricate reproducibly, and could hinder the scaling up of the number of qubits.

\subsection{Multiple-electron Semiconductor Qubits}

The main disadvantage of the single-electron spin-qubits discussed above  is that microwave signals are required for  single-qubit operations and that these operations are generally much slower than the two-qubit operations based on the exchange interaction. As an alternative, it has been proposed to also use this much faster exchange interaction for single-qubit rotations, but at the cost of requiring multiple electrons to implement one qubit.

\begin{table*}[ht]
\centering
\caption{\label{tab:specifications}Example specifications for a $\pi$-rotation on a single-electron spin qubit for a fidelity of 99.9\%.}\begin{tabular}{rl}
\hline
\textbf{Frequency inaccuracy}    & 11 kHz                                      \\
\textbf{Phase noise}             & -106 dBc/Hz at 1 MHz offset, -20 dB/dec slope \\
\textbf{Wideband additive noise} & 7.1 nV/$\sqrt{\text{Hz}}$                    \\
\textbf{Phase inaccuracy}        & 0.64 $^{\circ}$                               \\
\textbf{Amplitude inaccuracy}    & 14 $\mu$V on 2.0 mV amplitude                 \\
\textbf{Amplitude noise}         & SNR = -40 dB                                  \\
\textbf{Duration inaccuracy}     & 3.6 ns on 500 ns nominal duration            \\
\textbf{Timing jitter}           & 3.6 ns$_{rms}$                               \\
\hline
\end{tabular}
\end{table*}

In case of the singlet-triplet qubit, the singlet and triplet configurations of two electrons distributed over two quantum dots are used   as the basis states of the qubit \cite{levy2002, petta2005, foletti2009, maune2012, shulman2012, wu2014, nichol2017}. A coherence time of $T_2^*$ = 10 ns has been demonstrated in a GaAs sample, which can be significantly boosted with e.g.~strong magnetic field gradients ($T_2^*$ = 700 ns) \cite{nichol2017} or pumping schemes to obtain dynamic nuclear polarization \cite{foletti2009}. Single-qubit  rotations  that  flip  the  qubit  state (X rotations) are then implemented by suddenly pulsing the voltage on the gates close to the quantum dots (similar to two-qubit gates for single-electron spin-qubits), with typical $\pi$-rotations of $\sim$1 ns \cite{petta2005, shulman2012}. Single-qubit rotations altering the phase of the qubit (Z rotations) rely on an ever-present magnetic field gradient between the two quantum dots that continuously rotates the qubit. By timing the period between different operations, the desired rotation can be obtained. This rotation is generally slower, e.g.~10$\times$ slower in Ref. \cite{shulman2012}.
Two-qubit  gates can be implemented by capacitively coupling two pairs of quantum dots, each encoding a qubit, resulting in an operation time below \mbox{1 $\mu$s} \cite{shulman2012, nichol2017}. The strength of this capacitive coupling is approximately proportional to the exchange interaction in both qubits.
Read-out of the quantum state exploits the Pauli-exclusion principle  to convert the quantum state encoded in the spin into a different  charge occupation in the quantum dots, which is consecutively measured using a charge sensor.  

A drawback of singlet-triplet qubits is that a magnetic field gradient is required between the two quantum dots. The design can be simplified by adding a third electron to represent a single qubit, as is the case for the exchange-only qubit and the hybrid qubit. 
In the exchange-only qubit, three quantum dots in a row, each containing a single electron, are used to implement a single qubit \cite{divincenzo2000, gaudreau2012, medford2013, eng2015, shim2016, setiawan2014}. Using the pairwise exchange interaction between the middle quantum dot and one of the outer quantum dots, single qubit rotations can be obtained around two axes at a 120$^\circ$ angle. In Ref. \cite{eng2015}, exchange operations with a tunable duration from 0.01 $\mu$s to 1 $\mu$s have been measured around both axes, while obtaining a coherence time of $T_2^*$ = 2.3 $\mu$s in purified silicon. 
Proposals exist to implement two-qubit gates using only the exchange interactions between the quantum dots \cite{divincenzo2000, setiawan2014}.

In the hybrid qubit, the three electrons are placed in only two quantum dots, thereby simplifying the gate geometries and fabrication \cite{shi2012, koh2012, shi2014, ferraro2014, kim2014, kim2015}.
Single-qubit rotations around different axes are also obtained by suddenly pulsing the gates. In this way, $\pi$-rotations in less than \mbox{100 ps} have been demonstrated and coherence times of $T_2^*$ $\sim$ 10 ns have been achieved \cite{kim2014}.
For the two-qubit gates, it has been shown that in theory fewer exchange-interactions are required between two double quantum dots than in the case of the exchange-only qubit \cite{shi2012}.

\begin{figure*}[ht]
\centering
\includegraphics[width=0.7\linewidth]{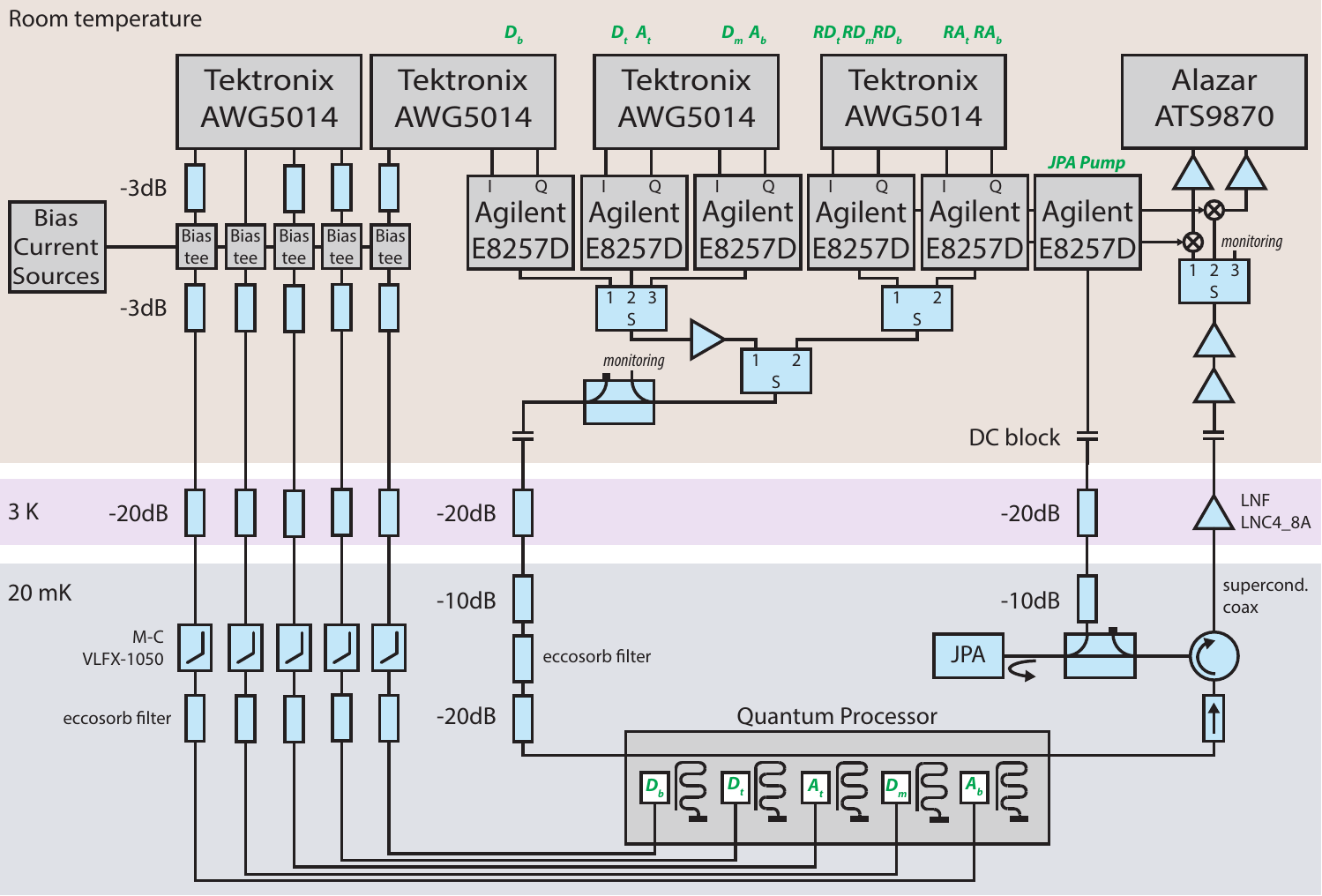}
\caption{\label{fig:setup_transmon}Example of a setup used for superconducting transmon qubits \cite{riste2015}.}\end{figure*}

\subsection{Controller specifications}
\label{sec:requirements_specs}

Table \ref{tab:technologies} summarizes the electrical signals  required for the different qubit technologies to drive the various gate operations and to be acquired for the read-out. 
In general, it is sufficient for the electronic controller to generate electrical voltage pulses in baseband, e.g.~for single and two-qubit gates in singlet-triplet qubits, or at microwave frequency, e.g.~for single-qubit gates for transmons. Likewise, the read-out electronics must demodulate pulses at baseband or at microwave frequencies. In addition, but not shown in Table \ref{tab:technologies}, the electronics must bias the quantum hardware appropriately, i.e.~provide stable voltage bias for all gates in spin-based systems and provide accurate current for the flux bias for transmons.

On top of providing the required functionalities, the properties of the driving signals  must be tightly controlled, both in terms of  accuracy in the  frequency, the amplitude and  the duration, and in terms of low electrical noise levels, in order not to degrade the inherent \emph{fidelity} of the qubits. The fidelity specifies how accurately an intended operation is performed. As the fidelity achieved in current qubit technologies is too low for algorithm execution, quantum error correction (QEC \cite{fowler2012surface}) can be used to improve the fidelity by encoding the quantum state on a larger number of physical qubits. QEC requires a minimum fidelity of the physical qubits and employs continuous cycles of qubit measurement with feedback control. Since feedback must be applied well within the qubit coherence timescale for QEC to be effective, fast qubit measurements and low controller latency are required. However, most state-of-the-art experiments do not apply QEC and feedback control, and, as a result, long measurement times ($< 1$ ms) are typically used to allow filtering of the noise and thereby increase the read-out fidelity in semiconductor qubits. For this reason, Table \ref{tab:technologies} does not report measurement times for semiconductor qubits, although QEC would ask for a read-out time in the order of 100 ns - 1 $\mu$s. The control fidelity is inherently limited by non-idealities in the quantum processor, but can be further limited by non-idealities in the applied control signals.

The effect of signal non-idealities can be either simulated \cite{van2018co} or analytically evaluated \cite{ball2016role, van2018impact}. As an example, the sources of inaccuracy and the related specifications of the  microwave pulse that achieves a 99.9\% fidelity for a $\pi$-rotation on a single-electron spin qubit are shown in Table \ref{tab:specifications}. These have been derived using the methods outlined in Ref. \cite{van2018impact}. Although a fidelity of 99.9\% has only recently been achieved in several qubit technologies, such fidelity level would be the absolute minimum for QEC \cite{fowler2012surface} thereby representing the absolute minimum specifications that should be targeted. It is important to note that the sensitivity to non-idealities in certain signal properties can be reduced by several techniques, such as pulse shaping \cite{khaneja2005optimal}, dynamical decoupling  \cite{uhrig2007keeping}, and by performing the qubit gates in a different operating point that is e.g.~less sensitive to charge noise \cite{martins2016noise}.

Furthermore, constraints  on the electrical interface are also set by qubit idling, i.e.~the periods during which no operation is performed on a qubit that must just preserve its quantum state. Examples are the noise level on the gates of an idle spin-based qubit and the crosstalk on an idle qubit while a microwave signal is driving the single-qubit rotation of a neighbouring qubit. However, the signal specifications are generally more relaxed during idle periods than when performing an operation. For example, the gate bias voltage  fixing the exchange interaction for spin qubits must be accurate  during the qubit operation, but during idling a lower accuracy can be tolerated, as long as the interaction is sufficiently low \cite{vandersypen2016}. Nonetheless, if the idle time is much longer than the operation time, specifications driven by idling can become significant. 

Surprisingly, the electrical requirements for each qubit technology have not yet been systematically analyzed, resulting in quantum-computing experimentalists to simply overdesign their electronic setups, as described in the following section. The interested reader is referred to Ref. \cite{van2018impact} for a complete analysis of the electrical specification for single-electron spin qubits, which shows a methodology that can in principle be adapted to any qubit technology.
 \begin{figure*}[ht]
\centering
\includegraphics[width=0.7\linewidth]{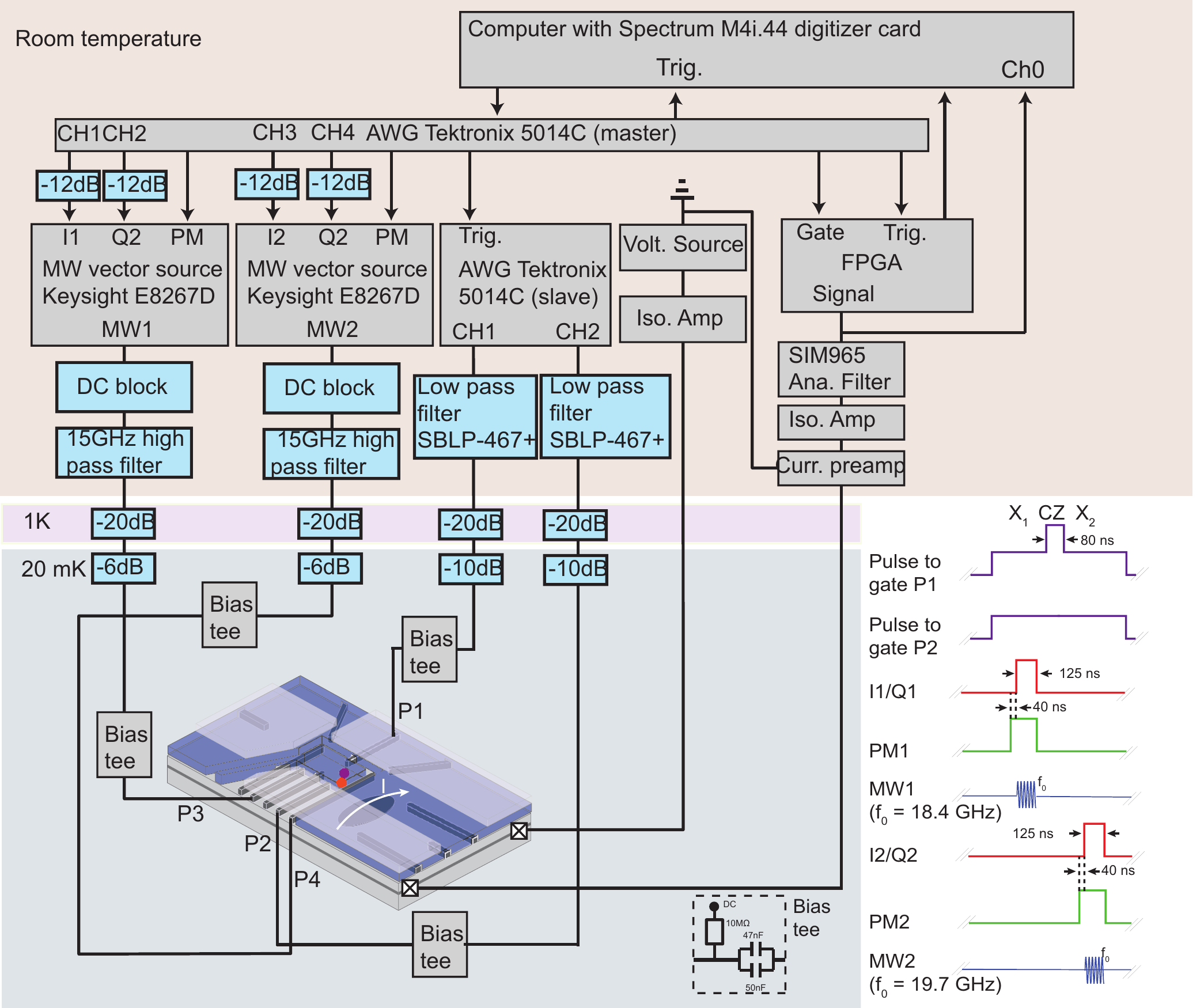}
\caption{\label{fig:setup_spin}Example of a setup used for single-electron spin qubits in quantum dots (image taken from Ref. \cite{watson2018}).}
\end{figure*}

\section{Review of State-of-the-Art Controllers\label{sec:controllers}}
In this section, we will show how state-of-the-art experimental setups generate the control signals and acquire the read-out signals, as required by existing quantum processors.

\subsection{Superconducting Qubits}
Fig. \ref{fig:setup_transmon} shows the experimental setup  used in Ref. \cite{riste2015} to control a five-qubit  transmon chip. A flux-bias line for each transmon is biased with a DC current to set the  frequency of the respective qubit. 

For two-qubit gates, pulses on the flux lines  are generated by an arbitrary waveform generator (AWG, here the 4-channel 1.2 GS/s Tektronix AWG5014), and added to the DC component via wideband 50-$\Omega$ bias-tees.  For  single-qubit gates, microwave signals need to be applied to the  input port of the respective cavity. The  microwave carrier from a vector signal generator (VSG, here the Agilent E8257D) can be I/Q modulated by an AWG to generate the required DRAG pulses, e.g.~for qubit $D_b$ in Fig. \ref{fig:setup_transmon}. Additionally,  the AWG can also provide a single-sideband modulation to drive a qubit at a frequency different from the carrier frequency provided by the VSG, as shown for all other qubits in Fig. \ref{fig:setup_transmon}.

The tones required for qubit read-out are generated in the same way, and all resulting carriers are combined before entering the cryogenic refrigerator. However, for the read-out, the carrier generated by the VSG is also fed into the acquisition circuit to down-convert the signal received from the cavity. A Josephson parametric amplifier (JPA \cite{mallet2009single}) at base temperature and a cryogenic low-noise amplifier (LNA \cite{LNA48}) are used to improve the signal-to-noise ratio (SNR) to allow for shorter measurement times. The pump signal required by the JPA is provided by a VSG. Additional gain stages are used at room temperature before down-converting the signal and digitizing it using a high-speed ADC card. Additional demodulation and signal processing is performed on a PC.

In addition to the components described above, several attenuators and filters are used at different temperature stages of the  refrigerator both to minimize the heat conducted through the cables to the cryogenic chamber and to reduce the noise.

\subsection{Semiconductor Qubits}
Due to the similarities in the different semiconductor qubit setups, we will only consider a setup for single-electron spin qubits in quantum dots, and note where other implementations differ. 
Fig. \ref{fig:setup_spin} shows the experimental setup  used in Ref. \cite{watson2018} to control a two-qubit single-electron spin-qubit chip.

Similar as for superconducting qubits, DC biasing of the quantum processor's various gates is required. While not shown in Fig.~\ref{fig:setup_spin}, all gates, including the top gates and the ones for the SET charge sensor, are connected to the proper DC voltage generator. 

The microwave signals required for the single-qubit gates are again generated by a VSG (the Keysight E8267D) which is modulated using an AWG (the Tektronix 5014C). Additional pulse modulation (PM) control is implemented to turn on the microwave carrier generation only when required to minimize signal leakage during idle periods.

The  voltage pulses on the gates, as required for qubit initialization, read-out and two-qubit gates, are provided by an AWG connected to the gate via a low-pass filter and a bias-tee. The pulse generation of this AWG is controlled using an external trigger provided by the master AWG that ensures proper synchronization.

For the multiple-electron spin qubits  that leverage the high operation speeds offered by the exchange interaction, the speed of an AWG might be insufficient to provide pulses that are short enough or have sufficient timing resolution for accurate operations. In that case, dedicated pulse generators can be used, e.g.~the Agilent 81134A pulse pattern generator \cite{kim2014}.

For the read-out, the impedance of either an SET (as in Fig. \ref{fig:setup_spin}) or a QPC is modulated by the qubit state, and can be directly read by measuring the device current when biased at a fixed voltage (Fig. \ref{fig:setup_spin}) \cite{vandersypen2004, vink2007}. The applied signal conditioning consists of (cryogenic) amplification and filtering, before the signal is digitized. Unfortunately, the bandwidth and hence the speed of this measurement is limited by the capacitive parasitics due to the wiring connecting the quantum device to the amplifiers. 
\begin{figure}[ht]
\centering
\includegraphics[width=0.8\linewidth]{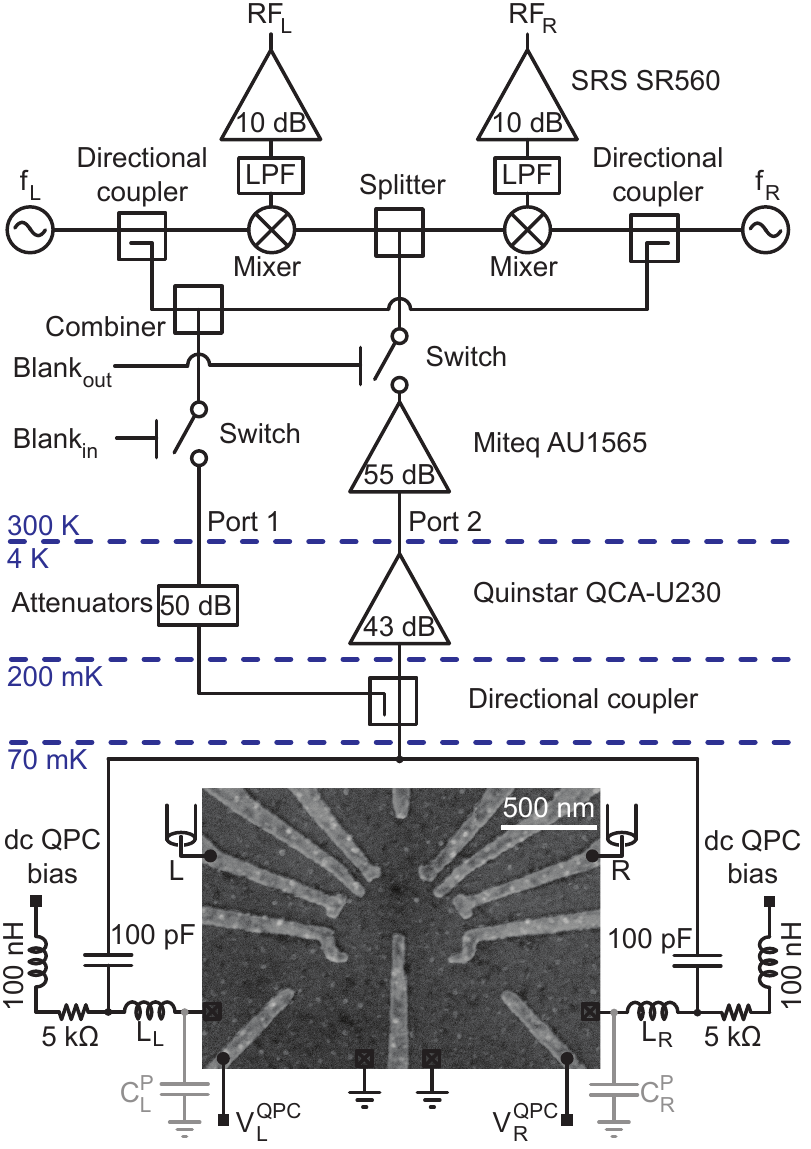}
\caption{\label{fig:setup_reflec}Example of a setup using RF-reflectometry of a QPC to sense the charge occupation of a triple quantum dot (image taken from Ref. \cite{laird2010}).}
\end{figure}
Alternatively, RF reflectometry is also commonly  used to increase the bandwidth of the   charge-sensor  impedance measurement \cite{reilly2007, hornibrook2014}. An example  using this method is shown in Fig. \ref{fig:setup_reflec} \cite{laird2010}. The nominal impedance of the charge sensor is adapted to 50 $\Omega$ by a matching network closely connected to the charge sensor. By sending an RF pulse and measuring the reflected power steered by a directional coupler, changes in the impedance can be monitored. Instead of using a charge sensor, dispersive read-out of the quantum dot's capacitance can be used to detect changes in the dot occupation, which uses a setup similar to the  RF-reflectometry read-out discussed above \cite{colless2013}.

\subsection{Controller Performance versus  Signal Requirements}
The use of general-purpose instruments, such as VSGs and AWGs, offers high flexibility, wide tuning ranges and rapid testing. However, to obtain accurate quantum operations as required to perform any quantum algorithm, the instrument specifications should be carefully evaluated. As a representative example, we will consider the generation of the microwave pulses specified in Table \ref{tab:specifications} and generated  in the setup of Fig. \ref{fig:setup_spin} by the Tektronix 5014C AWG \cite{Tektronix} and the Keysight E8267D VSG \cite{Keysight2016}. While a frequency inaccuracy of 11 kHz can be tolerated, the Keysight E8267D offers a 1 mHz frequency resolution, and its phase noise is more than 15 dB better than required. However, the instrument specifies a wideband additive noise floor of 63 nV/$\sqrt{\text{Hz}}$. Provided that more than 20-dB attenuation is used, this is also sufficient to meet the specification of 7.1~nV/$\sqrt{\text{Hz}}$. The VSG offers a phase offset resolution of 0.1$^{\circ}$, which is well within the specifications. The required amplitude accuracy specified in Table \ref{tab:specifications} translates into an 8-bit amplitude resolution, while the duration accuracy can be achieved with a sample rate of 150 MS/s. Those can be accommodated by the Tektronix 5014C that offers a 14-bit resolution with a 1.2 GS/s rate. The specified jitter of 5 ps$_{rms}$ is significantly better than the required 3.6 ns$_{rms}$.

In summary, the performance of the analyzed setup is sufficient to drive single-qubit operations in an ideal quantum processor with a 99.9\% fidelity. More in general, it can be concluded that typically adopted general-purpose instruments are not limiting the fidelity of state-of-the-art quantum computers, as shown in \cite{van2018impact},  since solid-state semiconductor qubits with fidelity exceeding 99.9\% have been only recently demonstrated \cite{yoneda201799}. While this situation is desirable for current developments focused on improving the performance of quantum processors, the large performance margin in the electronic interface may not be tolerated as the performance and the scale of quantum processors improve, as discussed in Section \ref{sec:mux}.
 \section{\label{sec:mux}Challenges in Scaling-up}

State-of-the-art controllers for quantum processors described in Section \ref{sec:controllers} are well suited  for current qubit experiments, but will show severe limitations  when scaling up to the thousands of qubits as required for any practical quantum computer.

First, the general-purpose instruments  used in current experiments are expensive, large and power hungry while only supporting a handful of qubits. Although this can be partially alleviated by sharing the hardware over multiple qubits, e.g.~as in Fig.~\ref{fig:setup_transmon} (more details in Section \ref{sec:muxoptions}), this can induce crosstalk effects between qubits, e.g.~undesired phase shifts on undriven qubits \cite{riste2015,van2018impact}. A more scalable solution is the adoption of tailor-made electronics ensuring the required specifications and optimized for size, power and cost. Multiple tailor-made controllers that have already been employed will be further discussed in Section \ref{sec:customcontrol}. Tailor-made controllers can also be optimized for speed, thus avoiding issues arising in feedback control due to  general-purpose instruments  generally being too slow. For instance, the Tektronix 5014C AWG (see Fig.~\ref{fig:setup_transmon}, Fig.~\ref{fig:setup_spin}) has a delay longer than 500 ns between the arrival of a trigger and the generation of the actual output signal \cite{Tektronix}.

\begin{table*}[htbp]
  \centering
  \caption{\label{tab:customcontrollers}Comparison of tailor-made room-temperature controllers.}
  \resizebox{\textwidth}{!}{    \begin{tabular}{l|llllll|}
    \textbf{Controller} & \textbf{Qin, 2017 \cite{qin2017}} & \textbf{Fu, 2017 \cite{fu2017}} & \textbf{Ryan, 2017 \cite{ryan2017}} & \textbf{Salathe, 2018 \cite{salathe2018}} & \textbf{Lin, 2018 \cite{lin2018high}} & \textbf{Ofek, 2016 \cite{ofek2016demonstrating}}\\
    \hline
    \textbf{FPGA} & Virtex-7 & Cyclone V & Virtex-6 & Virtex-4 & Xilinx & Virtex-6\\
    \textbf{Host interface } & USB 2.0 & USB   & 1 Gb Ethernet &  & 1 Gb Ethernet & PCIe \\
    \textbf{Peripherals} & 2 x TDC & 2 x 200 MS/s 8-bit ADC & 2 x 1 GS/s 12-bit ADC & 1 x 100 MS/s 14-bit ADC & & 2 x 1 GS/s ADC \\
          & 12 x pulse generator & 8 x marker outputs & 2 x 1 GS/s 16-bit DAC & & 2 x 2 GS/s 16-bit DAC & 2 x 1 GS/s DAC\\
    \textbf{Slave modules} & \textbf{1 DAC board} & \textbf{3 AWG modules} & \textbf{up to 9 APS2 modules} & &  & \\
    with each:     & 2 x 1 GS/s 16-bit DAC & 2 x 200 MS/s 14-bit DAC & 2 x 1.2 GS/s 14-bit DAC & & & \\
          &       &       & 4 x marker outputs & & & \\
    \textbf{Latency} & -     & 80 ns (AWG only) & 428 ns (+ 110 ns cabling) &  283 ns (+ 69 ns cabling) &  & 200 ns \\
    \hline
    \end{tabular}    }
\end{table*}

Another limitation is that, at the moment, every qubit requires individual control lines connecting the quantum processor in the  refrigerator to the instruments at room temperature, e.g.~the flux-bias lines for transmons and the gates for spin qubits. Since the number of lines that can be physically fitted in a dilution refrigerator is limited, the wiring can become a serious bottleneck in scaling-up. Moreover, more connections, especially between stages at different cryogenic temperatures, cause the system to become more complex, more expensive and  less reliable. Furthermore, the heat conducted through each line from the warmer stages to the colder ones,  together with the power dissipated in the various signal attenuators, adds to the thermal load of the dilution fridge, thus increasing the requirements on its cooling power. Lastly, because of the mere size of the typical dilution fridge and because of the large number of bulky instruments, the cables are long enough to cause delays that cannot be neglected in any case. For instance, a 1-meter cable  results in a round-trip time of $\sim$ 10 ns, which can be comparable to the duration of the quantum operations (Table \ref{tab:technologies}).

In order to limit the wiring complexity, two main approaches (or a combination of them) can be embraced \cite{levy2011, vandersypen2016}: multiplexing the control and read-out signals over a reduced set of wires (Section \ref{sec:muxoptions}) and/or move the electronic interface closer to the qubits and operate it at cryogenic temperatures (Section \ref{sec:cryocontrol}).
 \section{Scalable Electronic Controller\label{sec:scalable_controller}}
\subsection{\label{sec:customcontrol}Tailor-made room-temperature controllers}

Recently, various tailor-made quantum-processor controllers have been developed \cite{qin2017,fu2017,ryan2017,salathe2018,lin2018high,ofek2016demonstrating}, as  urged by the demand for highly integrated and flexible solutions for the rapidly evolving quantum processors. To maximize  flexibility, these solutions generally use a reconfigurable Field-Programmable Gate Array (FPGA) at their core. Low delay is of utmost importance for quantum algorithms requiring feedback control. Direct processing of the read-out and control signals in an FPGA has a significant advantage in terms of latency compared to any software solution, while keeping the flexibility of in-field programmability.  In addition, integrating the full controller around an FPGA eases the synchronization of the various components. A list of controllers and their capabilities is shown in Table \ref{tab:customcontrollers}. Additionally, commercial general-purpose instruments are evolving to fulfill the needs of quantum control, e.g.~the Quantum Researchers Toolkit by Keysight offers a scalable modular control platform with a latency below 150 ns \cite{keysightquantum} and Zurich Instruments' series of HDAWGs a latency below 50 ns \cite{ziquantum}.

All electronic interfaces employ an FPGA as the main controller, and include hardware for both the generation of signals and analog-to-digital converters (ADC) to digitize the read-out signals. Such hardware can be either tailor-made and assembled on separate slave boards or comprise general-purpose instruments as in the setups described in Section \ref{sec:controllers}. Processing of the read-out signals is performed in the FPGA. In general, this includes signal demodulation, filtering, decimation, signal integration and state discrimination. Special techniques can be used to bring down the latency, such as using frequency-selective kernels to combine the demodulation, filtering and integration in a single processing stage \cite{ryan2017}, or using digital mixing and multiplier-less filters \cite{salathe2018}.

On the control side, additional functionality can be added on top of straightforward digital-to-analog converters (DAC), such as  micro-operations \cite{fu2017}. The Arbitrary Pulse Sequencer (APS2) proposed in Ref. \cite{ryan2017} provides additional direct digital synthesis (DDS) to implement single-sideband modulation, thus   allowing multiple qubits to be controlled by a single VSG. To control the frequency offset and keep track of the qubit phase, multiple numerically controlled oscillators (NCOs) are implemented in the FPGAs.

One common feature of the controllers in Table \ref{tab:customcontrollers} is operation at room temperature, resulting in additional latency. Furthermore, these controllers still do not fully replace all required hardware for qubit interfacing, e.g.~microwave signal generation is not included.

 \subsection{Multiplexing solutions\label{sec:muxoptions}}
Multiplexing can help in reducing the number of wires, but comes with additional constraints  as summarized in this section for time-division and frequency-division multiplexing.

\paragraph{Time-division multiplexing}
In the simplest multiplexing form, the same line is alternately connected to different qubits by means of a switch (TDMA).  In the control architecture proposed in  Ref.~\cite{hornibrook2015},  calibrated control waveforms are applied to a single line (`prime-line') that is switched over multiple qubits depending on the addressing information  supplied by another set of lines (`address-line').  The proposed switch matrix is controlled by an FPGA operating at 4 K and placed at the base-temperature stage of the dilution fridge (20 mK). Switches can be implemented with either GaAs HEMTs or a proposed capacitive switch. In addition to the basic on/off switching, phase and amplitude modulation can also be implemented in the switch matrix by tuning the impedance of each switch \cite{hornibrook2015}. This allows for the generation of specific signals from a limited set of waveforms provided from room temperature, thus representing the simplest example of a cryogenic electronic controller. Similarly, a Vector Switch Matrix (VSM) operating at room temperature is proposed in Ref. \cite{asaad2016independent} to switch single-qubit-gate control pulses over multiple transmons. Additional electronics included with each switch enables the tuning of both the phase and amplitude so that the signals to each qubit can be individually calibrated to take into account qubit fabrication inhomogeneity. 
In order to further reduce the number of lines, the address can be sent over a serial interface \cite{levy2011}. 

In Ref. \cite{veldhorst2017silicon}, the addressing scheme is extended to a 2D array using word and bit lines as in classical memories. A crossbar addressing scheme, similar to the use of  word and bit lines, can theoretically also be directly used to control a large grid of qubits without the need for additional devices like switches, as proposed in Ref.~\cite{li2017crossbar, vandersypen2016}. While this is an effective method to reduce the amount of interconnects, it requires a certain level of homogeneity in the qubit grid to obtain sufficiently accurate operations.

\begin{figure}[htbp]
\centering
\includegraphics[width=0.8\linewidth]{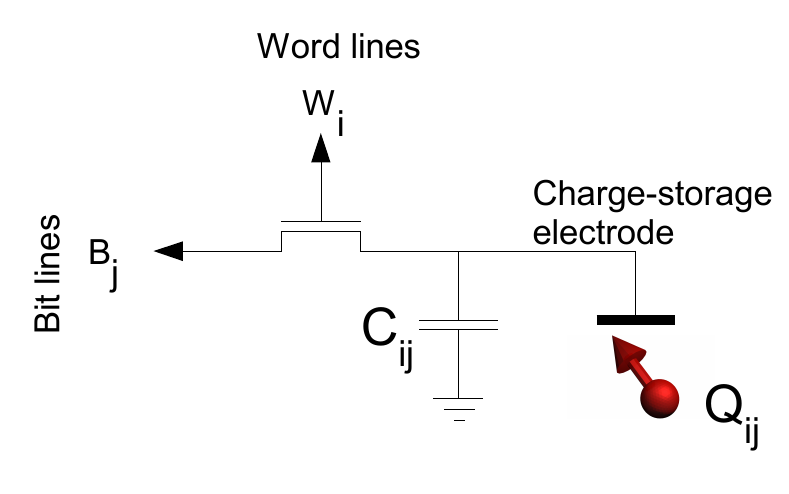}
\caption{\label{fig:tdma}Gate biasing using a local storage element and time division multiplexed access via word and bit lines (image taken from Ref. \cite{vandersypen2016}).}
\end{figure}

Time-division multiplexing may require the terminals of the quantum devices to remain floating for a given period. In order to keep such a terminal properly biased, as required e.g.~for gates in spin qubits, a sample-and-hold capacitor can be used to store the bias voltage at the gate  and must be only periodically refreshed, similar to a DRAM cell (Fig.~\ref{fig:tdma}) \cite{puddy2015multiplexed,schaal2018conditional,veldhorst2017silicon}. However, typically a relatively bulky capacitor in the order of 100 fF is required to achieve low enough voltage fluctuations, limited by either charge quantization or sampling noise \cite{vandersypen2016}.

While time-division multiplexing increases the scalability of the system, it strongly limits the amount of parallelism in the execution of quantum gates. As a result, not only the execution of the quantum algorithm becomes longer, but also the qubits will be idle for extended periods of time during which decoherence takes place, as discussed in details in Ref. \cite{levy2011}, thereby affecting the qubit fidelity. Furthermore, such multiplexing often requires a switch for each terminal of the quantum device. For qubits demanding a small pitch, such as spin qubits requiring $\sim$100 nm pitch between quantum dots, integrating even a single switching device, i.e.~a transistor, per qubit will be extremely challenging, as the needed density for the switching devices is higher than what is currently offered by the most advanced CMOS technologies \cite{wu20167nm}.

\paragraph{Frequency-division multiplexing}
In case of frequency-division multiple access (FDMA), different frequency bands on the same line are allocated for simultaneous use of different sub-blocks. As each individual superconducting transmon can be designed and tuned to  a different resonance frequency at which it is sensitive (Section \ref{sec:requirements}), frequency multiplexing can be directly used for control. In  Fig. \ref{fig:setup_transmon}, the read-out of different transmons also uses a different frequency band thanks to the use  of different read-out resonators tuned to different frequencies and coupled  to the same on-chip transmission line \cite{jerger2012frequency}.

Similarly, for single-qubit gates, single-electron spin qubits are only sensitive to signals at a given frequency, which is determined by the externally applied magnetic field. Frequency differences among neighbouring qubits, as required for FDMA, can be introduced using the Stark shift \cite{veldhorst2014, veldhorst2015} or by generating magnetic-field gradients by integrating on-chip micromagnets  \cite{kawakami2014,watson2018}. For an RF-reflectometry-based read-out, the use of FDMA is straightforward thanks to matching networks tuned to different frequencies \cite{hornibrook2014}, as shown in Fig. \ref{fig:setup_reflec}.

In order to limit the crosstalk between different channels, sufficient inter-channel frequency spacing  is required. In case of qubit control, the crosstalk causes the AC-Stark shift, i.e.~the qubit frequency appears to shift in the presence of signal power at a nearby frequency. As a result, either complex pulse shaping is required \cite{steffen2000simultaneous}, or the channels have to be time multiplexed with the addition of a simple phase correction for all idling qubits after each operation \cite{riste2015}. Because of practical limitations on the available spectrum, and the need for sufficient channel spacing, a maximum number of channels of only 10-100 is expected \cite{reilly2015}. Besides crosstalk, the application of frequency multiplexing in large-scale processors may be limited by the need for  large frequency-selective components, such as waveguide resonators for transmons  and matching networks in RF-reflectometry of spin qubits, although for the read-out this may be alleviated by crossbar approaches \cite{vandersypen2016}.
 \subsection{\label{sec:cryocontrol}Cryogenic Controllers}
\begin{figure}[tbp]
\centering
\includegraphics[width=0.9\linewidth]{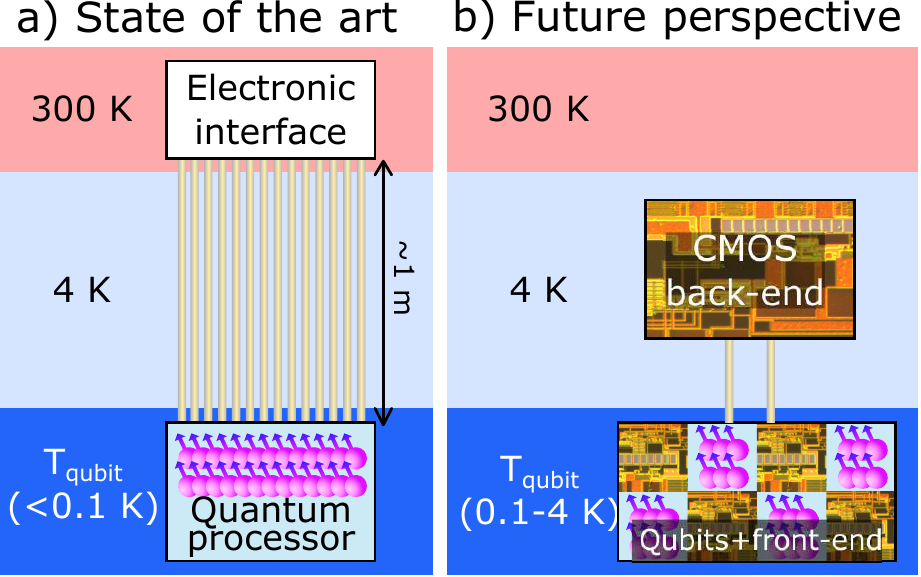}
\caption{\label{fig:cryocontroller}Implementing the controller inside the dilution refrigerator can significantly reduce the wiring complexity.}
\end{figure}

In order to further reduce the wiring complexity and latency, placing  the electronic controller close to the qubits, and hence operating it at cryogenic temperatures inside the dilution refrigerator, has been proposed \cite{sebastiano2017cryo, conway2016, reilly2015, ekanayake2007quantum, vandersypen2016, degenhardt2017, patra2018cryo}.
Ideally, the electronics should operate directly at base temperature next to the quantum processor. However, as the cooling power of typical refrigerators is lower at colder temperatures \cite{cryofridge}, only limited functionality could be implemented at sub-K temperatures. 
Some researchers anticipate that the power dissipation of the electronics can be reduced to a level compatible with the cooling power of existing refrigerators at sub-K temperatures, thanks to optimization of both the design and the microelectronic fabrication.
Although Ref. \cite{degenhardt2017} estimates that $\sim$100 singlet-triplet qubits can be controlled by electronics fabricated in a commercial 65-nm CMOS process operating at 100 mK, most researchers advocate for the largest section of the electronic interface to operate around  4 K (Fig. \ref{fig:cryocontroller}), i.e.~a temperature  at which  dilution refrigerators have significantly more cooling power (up to a few Watt). However, while a controller at 4 K solves the problems related to the wiring to room temperature, interfacing to a quantum processor operating at sub-K temperatures remains a challenge.
Relatively high-power electronics could be placed close to the quantum processor if thermally-isolating interconnect (such as superconducting through-silicon vias \cite{vahidpour2017superconducting}) could be employed to avoid heating up the qubits.
Using superconducting wires between the 4-K electronics and the quantum processor would drastically reduce the heat load on the sub-K stage \cite{tighe1999cryogenic, van2013niobium}, but would not alleviate complexity and reliability issues, which could otherwise be circumvented by multiplexing at base temperature (Section \ref{sec:muxoptions}).
Alternatively, preliminary studies promise the feasibility of some qubit technologies, spin qubits more specifically, operating at higher temperatures \cite{petit2018spin}. In that case, both  the quantum processor and its electronic interface could be placed at the same temperature (1-4 K) \cite{vandersypen2016}.

Several electronic technologies have proven their functionality down to deep cryogenic temperatures, and can hence be used to implement the controller. These include JFET, HEMT, superconducting devices based on Josephson junctions (e.g.~single flux quantum logic, SFQ \cite{katam2018}), compound semiconductors (e.g.~GaAs) and CMOS transistors \cite{cressler2013extreme,kirschman2014survey}. CMOS has shown functionality down to 30 mK \cite{ekanayake2010characterization, charbon2016cryo}, and it is the most advanced among these technologies. Thanks to the push of the  semiconductor industry, CMOS offers an established design automation infrastructure and  the possibility to integrate billions of transistors on a single chip, as would be required to interface millions of qubits. Consequently, it is generally considered the most promising choice for  the integration of a cryogenic electronic interface for quantum processors.

MOS transistors are fully functional at cryogenic temperatures, but their performance is different with respect to its standard temperature range. Improvements appearing at cryogenic temperatures include increased mobility, and hence larger maximum current, higher subthreshold slope, lower leakage and lower thermal noise \cite{incandela2017nanometer,homulle2018cryogenic}. As a drawback,  the threshold voltage increases, thus leading to less voltage headroom, and  flicker noise performance and device matching degrade \cite{dao2017impact, patra2018cryo}. As a major difference with respect to more mature CMOS technologies, advanced nanometer CMOS technologies are not affected by critical cryogenic non-idealities, such as current kink or hysteresis \cite{incandela2017nanometer,beckers2018characterization}. However, accurate device modelling is required even for nanometer CMOS technologies to enable the design of complex optimized cryogenic CMOS (cryo-CMOS) circuits \cite{incandela2017nanometer, sebastiano2017cryo}.

Nevertheless, some commercial CMOS integrated circuits (IC) are functional at cryogenic temperatures, well below their target temperature range. Most notably, some FPGAs remain fully functional down to 4 K with marginal change in operating speed  \cite{homulle2017reconfigurable}, and even DRAM seems functional down to 80 K \cite{tannu2017cryogenic}. The FPGAs could form the basis of a highly reconfigurable cryogenic control platform \cite{conway2016}, similar to the current tailor-made controller at room temperature (Section \ref{sec:customcontrol}). The ability to program the device in the field can prevent expensive and time-consuming warm-up/cool-down  cycles of the dilution fridge during development. In Ref. \cite{homulle2017reconfigurable}, the flexibility of a cryogenic FPGA platform has been demonstrated with the implementation of an FPGA-based reconfigurable ADC.

While some commercial ICs can be directly used out of the box, the best performance and highest level of integration can only be achieved by custom cryo-CMOS designs. Several components required for the control and read-out of a quantum processor have been designed and optimized for operation at cryogenic temperatures. In Ref. \cite{ekanayake2008},  pulse generators for qubit control using either a mixed-signal or a  fully-digital implementation have been integrated in a 500-nm Silicon-on-Sapphire (SOS) CMOS process. Both implementations can generate pulses with variable duration down to 10 ns. In Ref. \cite{rahman2017}, a 6-bit DAC designed in the same technology uses an analog calibration technique to overcome the increase in device mismatch at 4 K. A smaller 4-bit implementation demonstrates a fast rise time of 600 ps at 4.2 K \cite{rahman2017}. A 40-nm CMOS digitally controlled oscillator for signal generation at cryogenic temperature has demonstrated state-of-the-art phase-noise performance at 4 K while generating frequencies (5.5-7 GHz) compatible with  transmon control \cite{patra2018cryo}.

For the read-out, an LNA fabricated in a 500-nm SOS CMOS process shows  noise sufficiently low to measure the impedance of an SET in a measurement time of only $\sim$520 ns \cite{das2011low}. Another LNA, integrated  in a bulk 160-nm CMOS technology for the use in an RF-reflectometry read-out achieves a gain of 57 dB and bandwidth of 500 MHz at 4 K with an in-band noise figure of 0.1 dB (7 K noise temperature) \cite{patra2018cryo}. 

Besides these circuits, which are specifically designed for qubit control and read-out, various circuits operating at cryogenic temperatures exist for applications ranging from space missions to high-energy-physics, but are out of the scope of this review. Furthermore, while cryo-CMOS has many advantages, other technologies show superior performance in terms of noise (e.g. SiGe) or power consumption (e.g. SFQs) and could be a better alternative, e.g.~for the use in LNAs and FPGAs \cite{katam2018}, respectively. In fact, co-integration of a quantum processor and classical controller has already been demonstrated for an SFQ-based controller \cite{leonard2018digital}.

 \section{Conclusions\label{sec:conclusions}}

While state-of-the-art setups are well capable of fulfilling the requirements set by the different quantum technologies, they are subject to severe limitations when quantum processors will scale up. By using tailor-made electronic controllers at room temperature, the cost, size and power consumption can be reduced, while also enabling optimization for low-latency feedback. However, to obtain a truly scalable quantum computer, the interconnect complexity between the quantum processor and the electronic interface must be drastically reduced. While multiplexing techniques are effective in this sense, they still present a number of stringent drawbacks. On the contrary, the adoption of cryogenic electronics promises the deployment of truly scalable controllers. These cryogenic controllers can be implemented in standard CMOS technologies in order to exploit large-scale high-yield fabrication, thus allowing  the ultimate vision of co-integrating the electronic interface with the qubits on a single chip, or, at least, to operate the electronics in close proximity to the quantum processor. Initial steps have already been taken in this direction, thus paving the way to large-scale cryo-CMOS electronic interfaces that will make the operation of future quantum computers addressing real world-changing problems possible.

\section*{Acknowledgments}
The authors would like to thank Intel Corp. for funding this project.

\bibliographystyle{elsart-num}
\bibliography{main}

\end{document}